\documentclass[11pt]{article}

\usepackage[utf8]{inputenc}
\usepackage[T1]{fontenc}
\usepackage{amsmath,amssymb}
\usepackage{graphicx}
\usepackage{booktabs}
\usepackage{siunitx}
\usepackage{array}
\usepackage{adjustbox}
\usepackage{algorithm}
\usepackage{authblk}
\usepackage[margin=1in]{geometry}
\usepackage{xcolor}
\usepackage{caption}
\usepackage[numbers,sort&compress]{natbib}
\usepackage[colorlinks=true,linkcolor=blue,citecolor=blue,urlcolor=blue]{hyperref}

\title{\textbf{Constraint-Aware Quantum Optimization of Defect Configurations in Doped $ZrO_{2}$: XY-Mixer QAOA and Grover Adaptive Search}}

\author{Huajing Song\thanks{\texttt{wilsonsong85@gmail.com} \protect\newline \texttt{Huajing.Song@prattwhitney.com}}}
\affil{Materials \& Process Engineering Division, Pratt \& Whitney, An RTX Business, East Hartford, CT, USA}
\date{\today}

\begin{document}
\maketitle
% ===========================================================================
\begin{abstract}
% Lead with the validated FTQC oracle resource estimate; demote the
% full-vs-feasible ratio to a motivating observation.
Quantum optimization offers a route to searching the large defect-configuration spaces that arise in materials design. We develop an end-to-end, \emph{constraint-aware} quantum optimization workflow for composition-defect search in a Gd-doped ZrO\textsubscript{2} thermal-barrier-coating (TBC) material system, using a MACE-MPA-0 energy dataset to fit a 24-variable QUBO over 8 cation-occupation and 16 oxygen-vacancy variables with exactly two rare-earth substitutions and one oxygen vacancy, yielding 448 feasible configurations. The QUBO surrogate reproduces the MACE energies with held-out $R^2=0.997$ (full-data $R^2=0.999$, RMSE $=\SI{17}{\milli\electronvolt}$). We validate two complementary quantum pathways against exact enumeration: a constraint-preserving XY-mixer QAOA that confines sampling to the feasible subspace and places \SI{86}{\percent} of probability mass within \SI{1}{\milli\electronvolt} of the MACE optimum at depth $p=3$, and a fault-tolerant constrained Grover Adaptive Search oracle with explicit fixed-point arithmetic, branch-safe comparison, feasibility checking, and phase kickback. Across threshold cases, the validated oracle uses 324 high-level logical qubits, or 352--358 with conservative clean-ancilla v-chain accounting, and requires $3.6\text{--}4.3\times10^{4}$ Toffoli gates per Grover/GAS iteration. An idealized feasible-space amplification estimate suggests up to a $\sim240\times$ reduction in total Toffoli cost relative to the full $2^{24}$ occupation space, providing a resource-estimation bridge between materials-informed QUBO modeling, constraint-aware QAOA, and fault-tolerant threshold search.
\end{abstract}

% ===========================================================================
\section{Introduction}

Defect-configuration optimization is a central challenge in materials design:
the number of substitutional and vacancy arrangements grows combinatorially
with supercell size and chemical complexity. This challenge is especially
important in zirconia-based thermal barrier materials, where aliovalent rare-earth (RE) doping stabilizes high-temperature phases, introduces oxygen vacancies, and modifies thermal transport, sintering resistance, and mechanical durability~\citep{Padture2002TBC,Clarke2003TBC,Wu2002RareEarthZirconates}.
Yttria-stabilized zirconia (YSZ) remains the benchmark thermal-barrier-coating (TBC) topcoat, but next-generation rare-earth-modified zirconia and zirconate
chemistries are actively explored because cation size, dopant distribution, and
oxygen-vacancy ordering can strongly affect phase stability and thermal
conductivity~\citep{Clarke2003TBC,Wu2002RareEarthZirconates,Dong2017YSZLandscape}. In $RE$-doped ZrO\textsubscript{2}, the relative arrangement of cation
substitutions and oxygen vacancies therefore defines a discrete, constrained
energy landscape whose low-energy configurations are relevant to subsequent
transport and stability analysis.

Machine-learned interatomic potentials (MLIPs), including MACE and its materials
foundation-model variants, make it practical to evaluate large numbers of
candidate atomic configurations at substantially lower cost than direct
first-principles enumeration~\citep{Batatia2022MACE,Batatia2025Foundation}.
However, converting such energy data into a quantum optimization workflow
requires more than an energy predictor: the configurational objective must be
encoded as a binary optimization model, its constraints must be handled
correctly, and the resulting quantum cost Hamiltonian or oracle must be
validated against an exact classical reference. We therefore formulate the
fixed-composition defect-search problem as a quadratic unconstrained binary
optimization (QUBO) model, a standard representation connecting discrete
optimization, Ising Hamiltonians, and quantum optimization
algorithms~\citep{Lucas2014,Glover2022QUBO,Hucht_2011AnisotropySmallClusters}.

\begin{figure}[t]
\centering
\includegraphics[width=0.95\linewidth]{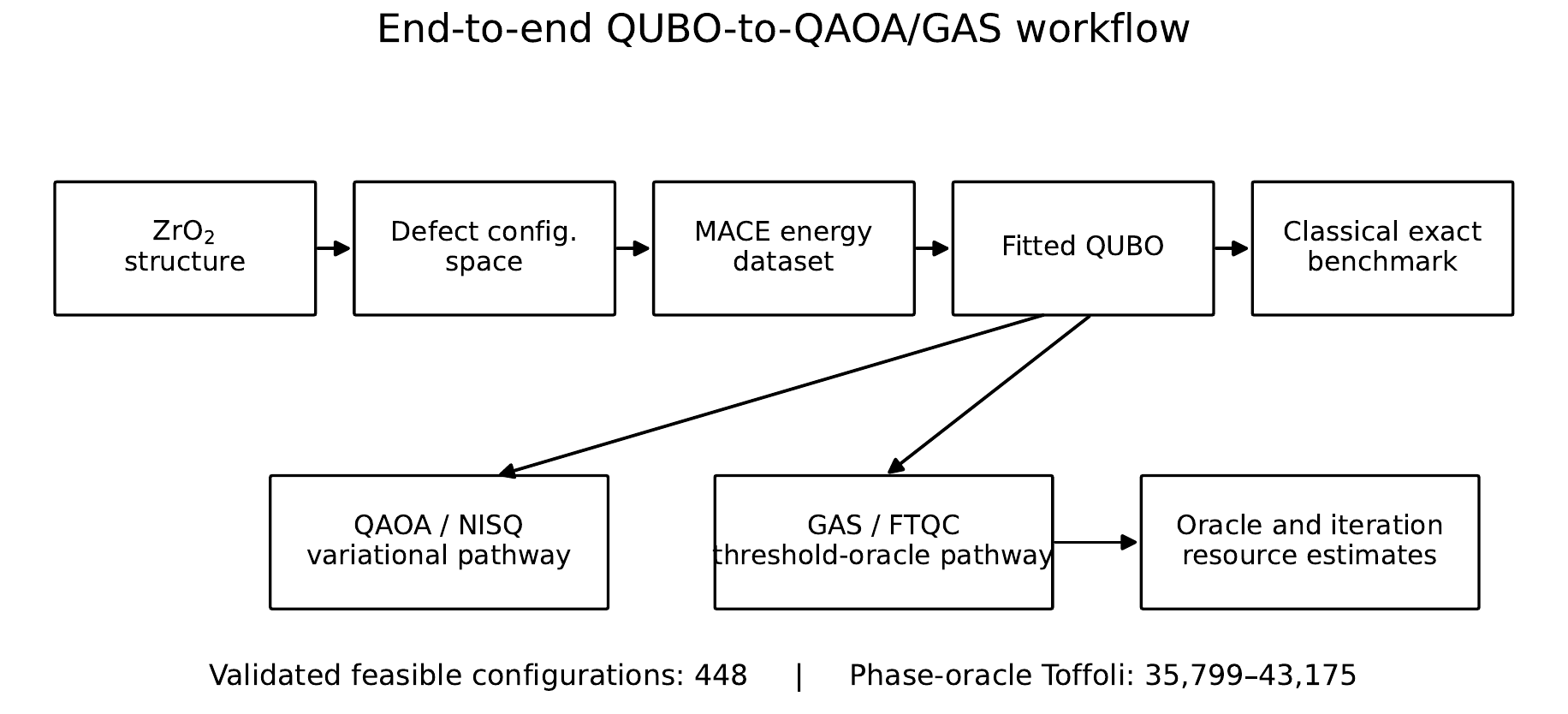}
\caption{End-to-end constraint-aware workflow. Starting from a
ZrO\textsubscript{2} supercell and its fixed-composition defect-configuration
space, MACE-MPA-0 energies are fitted to a QUBO and validated by exact
classical enumeration. The same objective is then optimized along two
complementary quantum pathways: constraint-preserving XY-mixer QAOA for the
near-term variational setting and a constrained Grover Adaptive Search oracle
for the fault-tolerant setting.}
\label{fig:workflow}
\end{figure}

The central message of this work is that \emph{constraint awareness} is the
decisive design choice on both sides of the quantum optimization stack. Using the Gd-doped ZrO\textsubscript{2} TBC material system as an example, we map a fixed-composition Gd-substitution defect problem to a QUBO and study
two complementary quantum pathways (Fig.~\ref{fig:workflow}): a near-term variational route based on the Quantum Approximate Optimization Algorithm (QAOA)~\citep{Farhi2014QAOA} and a fault-tolerant threshold-search route based on Grover Adaptive Search (GAS)~\citep{Gilliam2021GAS}. On the variational side, we use a constraint-preserving XY mixer within the quantum alternating-operator
framework~\citep{Hadfield2019QAOA,Fuchs2022ConstrainedMixers}, which confines
the dynamics to the fixed-composition feasible subspace instead of relying on
penalty terms to discourage infeasible samples. On the fault-tolerant side, we
explicitly construct and validate the reversible arithmetic required by a
constrained GAS oracle and quantify its logical-qubit and Toffoli costs. Exact
classical enumeration over the 448 feasible configurations provides the common
validation bridge between the surrogate model, the QAOA benchmarks, and the
fault-tolerant oracle construction.

A central aim of this paper is therefore to go beyond a black-box use of a
high-level Grover optimizer. We expose the logical-resource requirements of
applying GAS to a materials-informed QUBO by constructing and validating, layer
by layer, the fixed-point arithmetic, weighted accumulation, branch-safe
comparison, feasibility checking, bit-oracle, and phase-oracle stages. The main
contributions are:

\begin{enumerate}
\item A fixed-composition Gd\textsuperscript{3+} defect problem is mapped to a
24-variable QUBO with 448 feasible configurations, fitted to MACE-MPA-0~\citep{Batatia2025Foundation}
energies with held-out $R^2 = 0.997$ and validated by exact enumeration
(Sec.~\ref{sec:qubo}).

\item A constraint-preserving XY-mixer QAOA is benchmarked against penalty QAOA
on the same objective, including logical-resource and first-order noise
analysis, demonstrating the value of constraint awareness for near-term
optimization (Sec.~\ref{sec:qaoa}).

\item A constrained GAS phase oracle is constructed from fixed-point reversible
arithmetic, validated as reloadable circuit artifacts, and resource-estimated
in logical qubits and Toffoli gates
(Secs.~\ref{sec:gas}--\ref{sec:oracle}).

\item A one-iteration Grover/GAS baseline is reported, and the contrast between
the full occupation space and the feasible subspace is used to motivate
constraint-aware amplitude amplification as a clearly stated upper bound rather
than an achieved implementation (Sec.~\ref{sec:iteration}).
\end{enumerate}

% ===========================================================================
\section{Materials defect-configuration problem}
\label{sec:problem}

\begin{figure}[ht]
\centering
\includegraphics[width=0.8\linewidth]{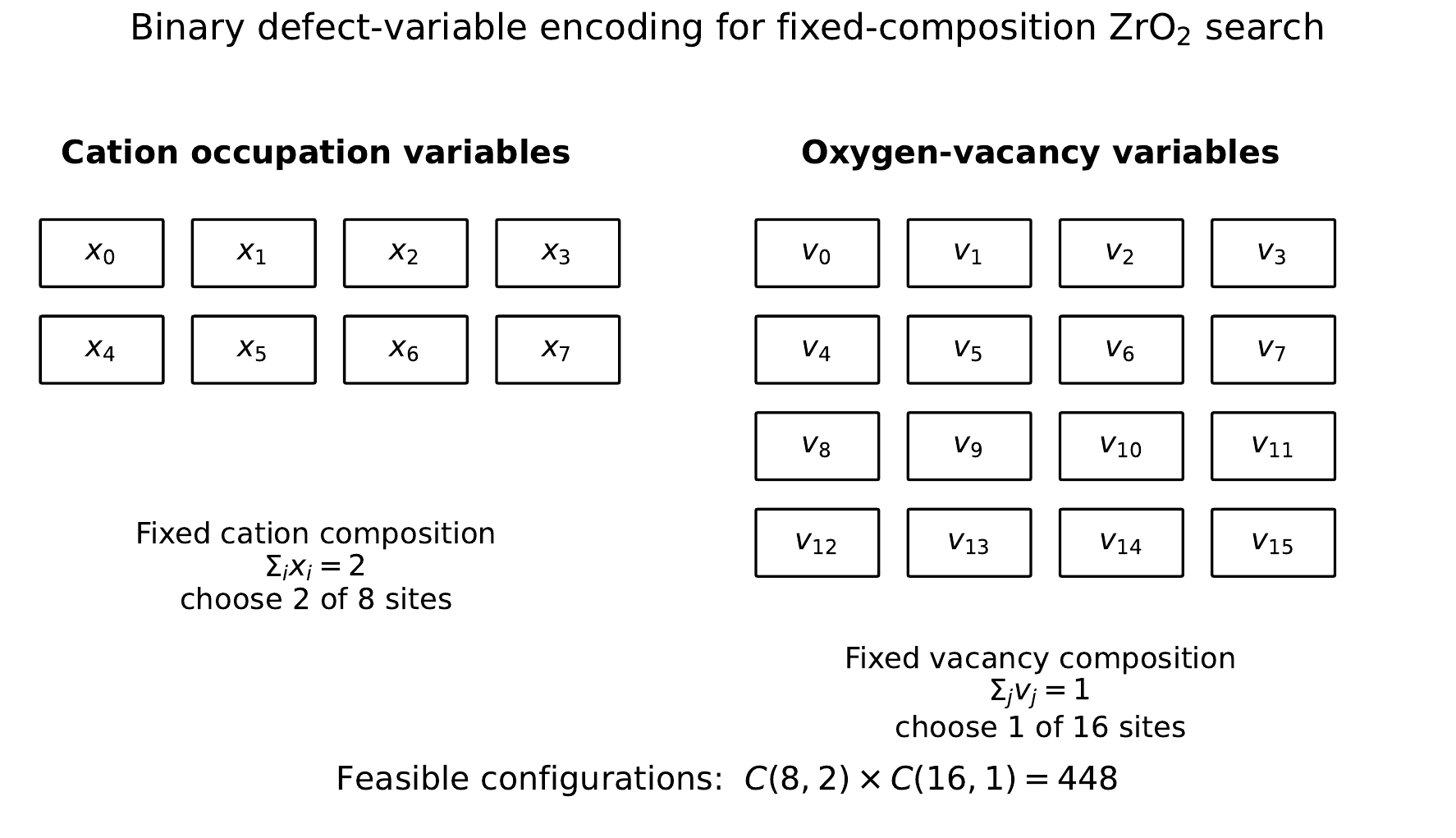}
\caption{Binary occupation encoding for the fixed-composition ZrO\textsubscript{2} defect search. Cation variables represent rare-earth substitution choices; anion variables represent oxygen-vacancy choices. The constraint $\sum_i x_i=2$ fixes the number of Gd substitutions, while $\sum_j v_j=1$ fixes the number of oxygen vacancies.}
\label{fig:encoding}
\end{figure}

We consider a $2{\times}2{\times}2$ fluorite ZrO\textsubscript{2} supercell
containing 32 cation sites and 64 anion sites. The cubic simulation cell has side length $L=\SI{10.18029082}{\angstrom}$, corresponding to a conventional fluorite lattice parameter of $a_0=L/2=\SI{5.09014541}{\angstrom}$. Within this supercell, we define a local defect block containing 8 cation sites and 16 anion sites and assign binary occupation variables to this block. The cation variables $x_i \in {0,1}$ encode rare-earth substitution by Gd on cation site $i$, with $x_i=1$ indicating Gd and $x_i=0$ indicating Zr. The anion variables $v_j \in {0,1}$ encode oxygen-vacancy placement on anion site $j$, with $v_j=1$ indicating a vacancy and $v_j=0$ indicating an occupied oxygen site. The fixed-composition constraints are
\begin{equation}
\sum_{i=1}^{8} x_i = 2, \qquad \sum_{j=1}^{16} v_j = 1 ,
\end{equation}
corresponding to exactly two Gd substitutions and one oxygen vacancy in the active block. The number of feasible configurations is therefore \begin{equation} \binom{8}{2}\binom{16}{1} = 28 \times 16 = 448 , \end{equation} embedded in a full 24-bit occupation space of $2^{24}=\num{16777216}$ bitstrings. This problem size is deliberately chosen: it is small enough to permit exact classical enumeration, while already capturing the essential structure of a materials defect-optimization problem, namely coupled substitution--vacancy energetics, fixed-composition constraints, and a binary occupation representation (Fig.~\ref{fig:encoding}) suitable for QUBO and Ising-type quantum optimization formulations~\citep{Lucas2014,Glover2022QUBO}.

% ===========================================================================
\section{MACE-derived QUBO and classical benchmark}
\label{sec:qubo}

\paragraph{Energy data.} For each of the 448 feasible configurations, the selected cation sites were assigned as Gd substitutions and the selected anion site was removed to create one oxygen vacancy. Energies were evaluated using the MACE-MPA-0 foundation model~\citep{Batatia2025Foundation}, medium variant, on the Gd-doped ZrO\textsubscript{2} supercell defined above. The calculations used the fixed cubic lattice vectors with $L=\SI{10.18029082}{\angstrom}$ and the corresponding ideal fluorite-site coordinates. No configuration-specific ionic or cell relaxation was performed before fitting the QUBO; the dataset therefore represents MACE single-point configurational energies on a common fixed geometry. This protocol isolates the discrete occupation dependence of the energy and provides a controlled benchmark for quantum optimization. In a materials-design workflow, the low-energy configurations sampled by the quantum algorithm can subsequently be re-ranked or refined using relaxed MACE or first-principles calculations.

\begin{figure}[h]
\centering
\includegraphics[width=0.7\linewidth]{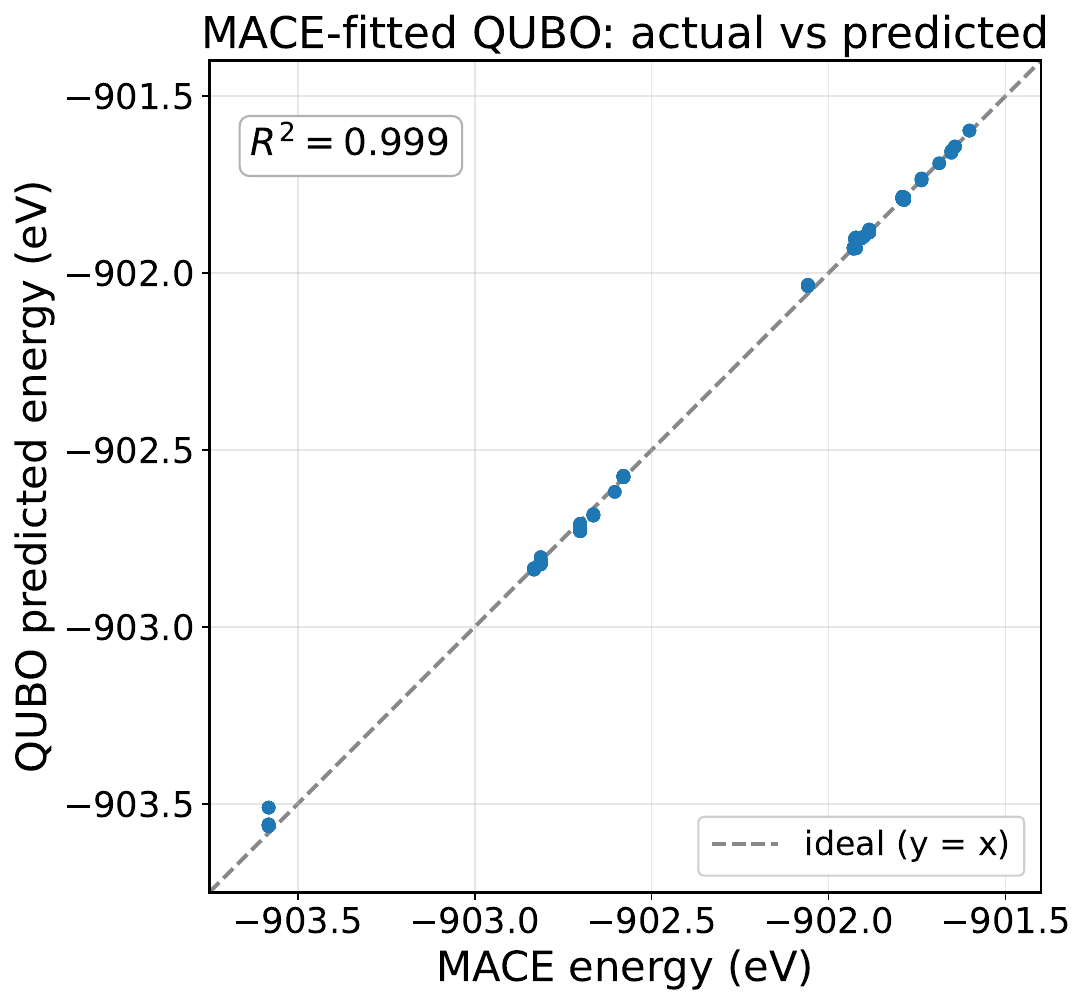}
\caption{QUBO-predicted versus MACE single-point energies for all 448 feasible configurations. The fitted quadratic surrogate reproduces the MACE configurational energy landscape with full-data $R^2=0.9991$.}
\label{fig:fit}
\end{figure}

\paragraph{QUBO fit.} The fitted QUBO can be viewed both as a quantum-optimization objective and as a pair-truncated occupation-variable surrogate for the configurational energy, closely related in spirit to cluster-expansion models used in alloy and defect thermodynamics~\citep{Sanchez1984ClusterExpansion,VanDeWalle2002ClusterExpansion}. We fit a quadratic surrogate of the form
\begin{equation}
E_{\mathrm{QUBO}}(x,v) = E_0 + \sum_i h_i^{c} x_i + \sum_j h_j^{v} v_j
+ \sum_{i<k} J_{ik}^{cc} x_i x_k + \sum_{i,j} J_{ij}^{cv} x_i v_j ,
\end{equation}
where the linear terms describe site-dependent substitution and vacancy
contributions, the cation--cation terms describe pairwise Gd--Gd interactions
within the active block, and the cation--vacancy terms describe coupled
substitution--vacancy energetics. Vacancy--vacancy terms are omitted because
the fixed-vacancy constraint $\sum_j v_j=1$ implies that $v_jv_l=0$ for all
distinct vacancy sites $j\ne l$ within the feasible subspace. Ridge regression,
a standard regularized least-squares method for stabilizing fits with
correlated or nonorthogonal predictors~\citep{Hoerl1970Ridge}, was used with
regularization $\alpha=10^{-3}$ to fit the surrogate over 180 features,
consisting of 24 linear terms and 156 quadratic terms. The resulting QUBO reproduces the MACE single-point energies accurately in aggregate. The full-data coefficient of determination is $R^2=0.9991$, with RMSE $=\SI{16.9}{\milli\electronvolt}$, MAE $=\SI{11.5}{\milli\electronvolt}$, and maximum absolute error $\SI{71.9}{\milli\electronvolt}$. A held-out test split gives $R^2=0.9974$ and RMSE $=\SI{27.6}{\milli\electronvolt}$ (Table~\ref{tab:fit}, Fig.~\ref{fig:fit}). These results indicate that a quadratic binary model captures the dominant configurational energy variation over the fixed-composition defect space.

\begin{table}[h]
\centering
\caption{QUBO-to-MACE fit fidelity. Energy regression is excellent in
aggregate; top-$k$ ranking overlap improves with $k$ as the near-degenerate
low-energy manifold is resolved.}
\label{tab:fit}
\begin{tabular}{lc}
\toprule
Metric & Value \\
\midrule
Features (linear + quadratic) & $24 + 156 = 180$ \\
Ridge regularization $\alpha$ & $10^{-3}$ \\
Full-data $R^2$ & $0.9991$ \\
Full-data RMSE / MAE (eV) & $0.0169$ / $0.0115$ \\
Full-data max.\ abs.\ error (eV) & $0.0719$ \\
Held-out test $R^2$ & $0.9974$ \\
Held-out test RMSE (eV) & $0.0276$ \\
\midrule
Top-5 ranking overlap & $0.40$ \\
Top-10 ranking overlap & $0.50$ \\
Top-20 ranking overlap & $0.75$ \\
Top-50 ranking overlap & $0.96$ \\
Top-100 ranking overlap & $0.90$ \\
\bottomrule
\end{tabular}
\end{table}

\paragraph{Ranking fidelity and low-energy degeneracy.} High aggregate regression accuracy does not guarantee perfect ordering of the lowest-energy configurations, because the low-energy manifold is near-degenerate. In this dataset, the QUBO global minimum (configuration~11) is the second-lowest configuration under MACE, while the true MACE minimum (configuration~203) lies only \SI{4.8}{\milli\electronvolt} above the QUBO minimum in QUBO energy. The top-$k$ overlap between the MACE and QUBO rankings increases from $0.40$ at $k=5$ to $0.96$ at $k=50$ (Table~\ref{tab:fit}). We therefore treat the QUBO as a faithful \emph{energy-landscape surrogate} for quantum-algorithm benchmarking, while explicitly recognizing that exact identification of the MACE ground state may require higher-order terms, local relaxation, or a final MACE re-ranking of quantum-sampled candidates. This hybrid re-ranking step is particularly natural for QAOA, where the output is a distribution over candidate configurations rather than a single deterministic structure.

\begin{figure}[h]
\centering
\includegraphics[width=0.95\linewidth]{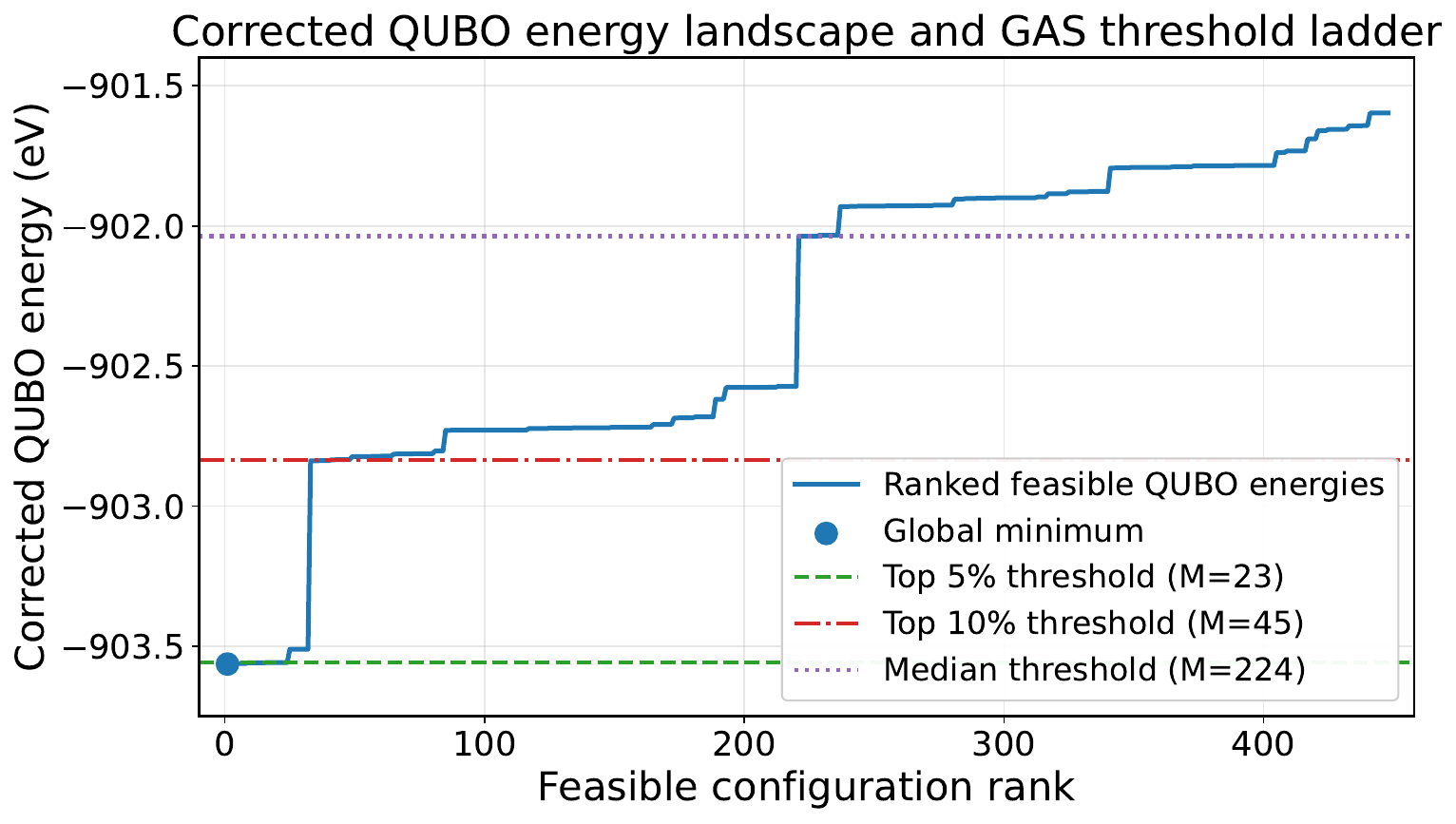}
\caption{Ranked corrected-QUBO energies over the 448 feasible configurations,
annotated with the threshold cases used as adaptive-GAS snapshots
(Sec.~\ref{sec:gas}).}
\label{fig:landscape}
\end{figure}

\paragraph{Classical benchmark.}
Exact enumeration over the 448 feasible configurations provides the classical reference for all subsequent quantum-algorithm tests. Enumeration identifies the QUBO and MACE low-energy configurations (Fig.~\ref{fig:landscape}), quantifies ranking agreement, and supplies exact marked-state counts for each Grover Adaptive Search threshold case used in Sec.~\ref{sec:gas}. This exact benchmark is essential because it allows the near-term QAOA results and the fault-tolerant GAS oracle validation to be checked against the same discrete objective and feasible subspace.

% ===========================================================================
\section{Constraint-aware variational optimization (QAOA)}
\label{sec:qaoa}

The fitted QUBO defines a diagonal cost Hamiltonian in the 24-qubit occupation basis and can therefore be used directly as the objective for the Quantum Approximate Optimization Algorithm (QAOA)~\citep{Farhi2014QAOA}. At depth $p$, the variational state has the form
\begin{equation}
|\psi_p(\boldsymbol{\gamma},\boldsymbol{\beta})\rangle =
\prod_{\ell=1}^{p}
e^{-i\beta_\ell H_M}
e^{-i\gamma_\ell H_C}
|\psi_0\rangle ,
\end{equation}
where $H_C$ is the QUBO-derived cost Hamiltonian and $H_M$ is the mixer. The fixed-composition constraints can be handled in two ways: by adding penalty terms to the cost Hamiltonian and searching the full $2^{24}$ occupation space, or by choosing a mixer that preserves the feasible subspace by construction. We benchmark both approaches in the occupation representation and validate all sampled configurations against exact enumeration and the underlying MACE energies.

\begin{figure}[h]
\centering
\includegraphics[width=0.72\linewidth]{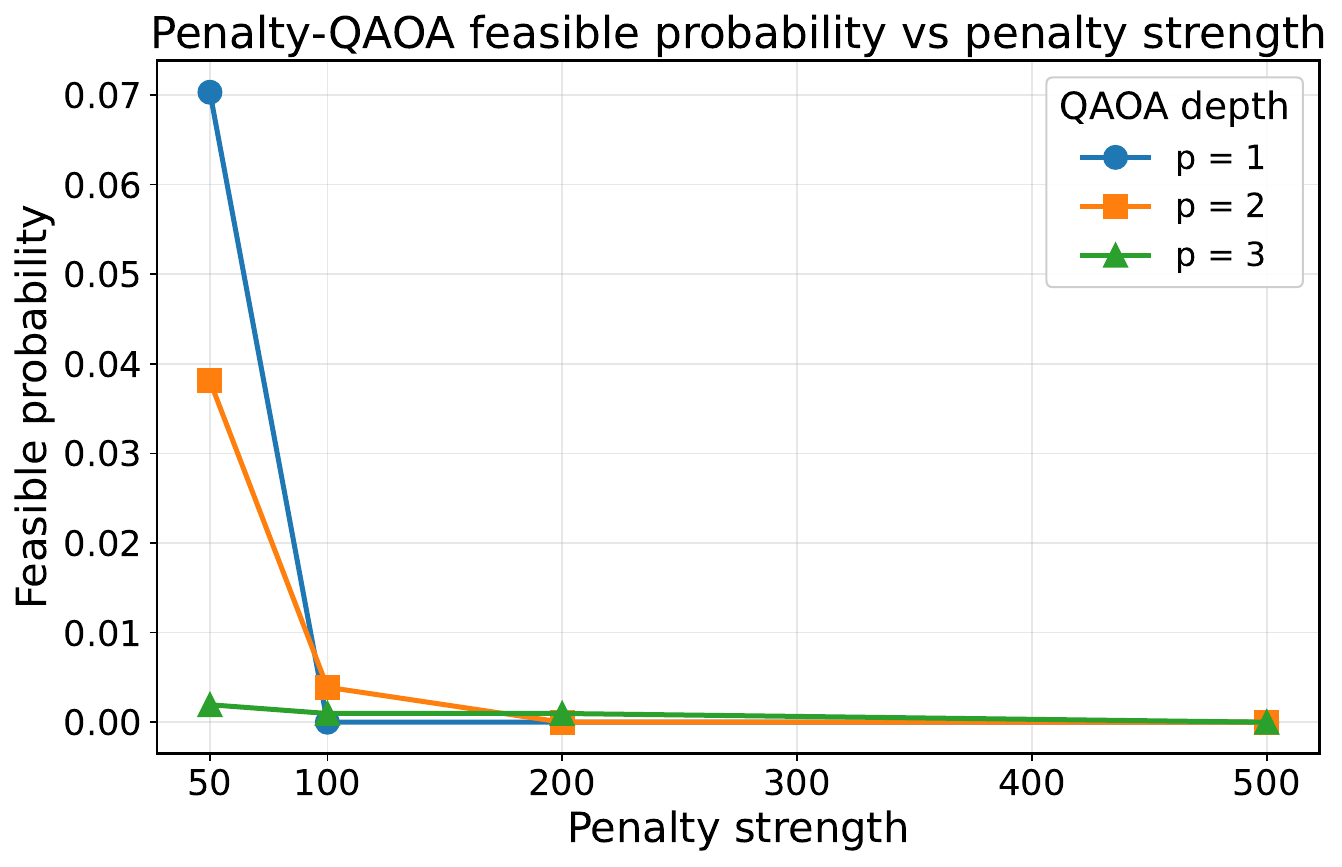}
\caption{Penalty QAOA feasible probability as a function of penalty strength.
Feasibility is low and collapses to zero for strong penalties, motivating a
constraint-preserving mixer.}
\label{fig:penalty}
\end{figure}

\paragraph{Penalty QAOA struggles with feasibility.} In the penalty formulation, the unconstrained cost is augmented by quadratic constraint penalties,
\begin{equation}
H_C^{\mathrm{pen}} = H_C
+\lambda_c\left(\sum_{i=1}^{8} x_i - 2\right)^2
+\lambda_v\left(\sum_{j=1}^{16} v_j - 1\right)^2 ,
\end{equation}
and the standard transverse-field mixer explores the full Hilbert space. This
approach is simple, but fragile for the present fixed-composition problem:
penalties must be large enough to suppress infeasible states, yet not so large
that they dominate the energy landscape and degrade optimization. Across 12
tested settings, corresponding to depths $p\in\{1,2,3\}$ and penalty strengths
$\lambda\in\{50,100,200,500\}$, only 6 produced any feasible sample. The best
feasible probability was \SI{7}{\percent}, and stronger penalties drove the
feasible probability to zero (Table~\ref{tab:qaoa}, Fig.~\ref{fig:penalty}).
The best penalty case that was both feasible and low in energy reached the QUBO ground-state energy to within \SI{0.1}{\milli\electronvolt}, but assigned only \SI{3.8}{\percent} probability to feasible samples. Thus, the penalty encoding does not provide reliable feasible-state preparation for this materials QUBO.

\paragraph{XY-mixer QAOA enforces constraints by construction.} The constraint-preserving alternative follows the quantum alternating-operator framework~\citep{Hadfield2019QAOA} and replaces the transverse-field mixer with fixed-Hamming-weight XY mixers acting separately on the cation and vacancy sectors~\citep{Fuchs2022ConstrainedMixers}. The mixer has the form
\begin{equation}
H_M^{\mathrm{XY}} = \sum_{(i,k)\in \mathcal{E}_c} \left(X_iX_k + Y_iY_k\right)
+ \sum_{(j,l)\in \mathcal{E}_v} \left(X_jX_l + Y_jY_l\right),
\end{equation}
where $\mathcal{E}_c$ and $\mathcal{E}_v$ are mixer graphs over the 8 cation and 16 vacancy variables, respectively. Because XY exchange conserves Hamming weight within each sector, initializing the circuit in any feasible state with $\sum_i x_i=2$ and $\sum_j v_j=1$ confines the entire QAOA evolution to the 448-dimensional feasible subspace. Feasibility is therefore exactly one by construction, independent of depth or variational parameters.

\begin{figure}[h]
\centering
\includegraphics[width=0.72\linewidth]{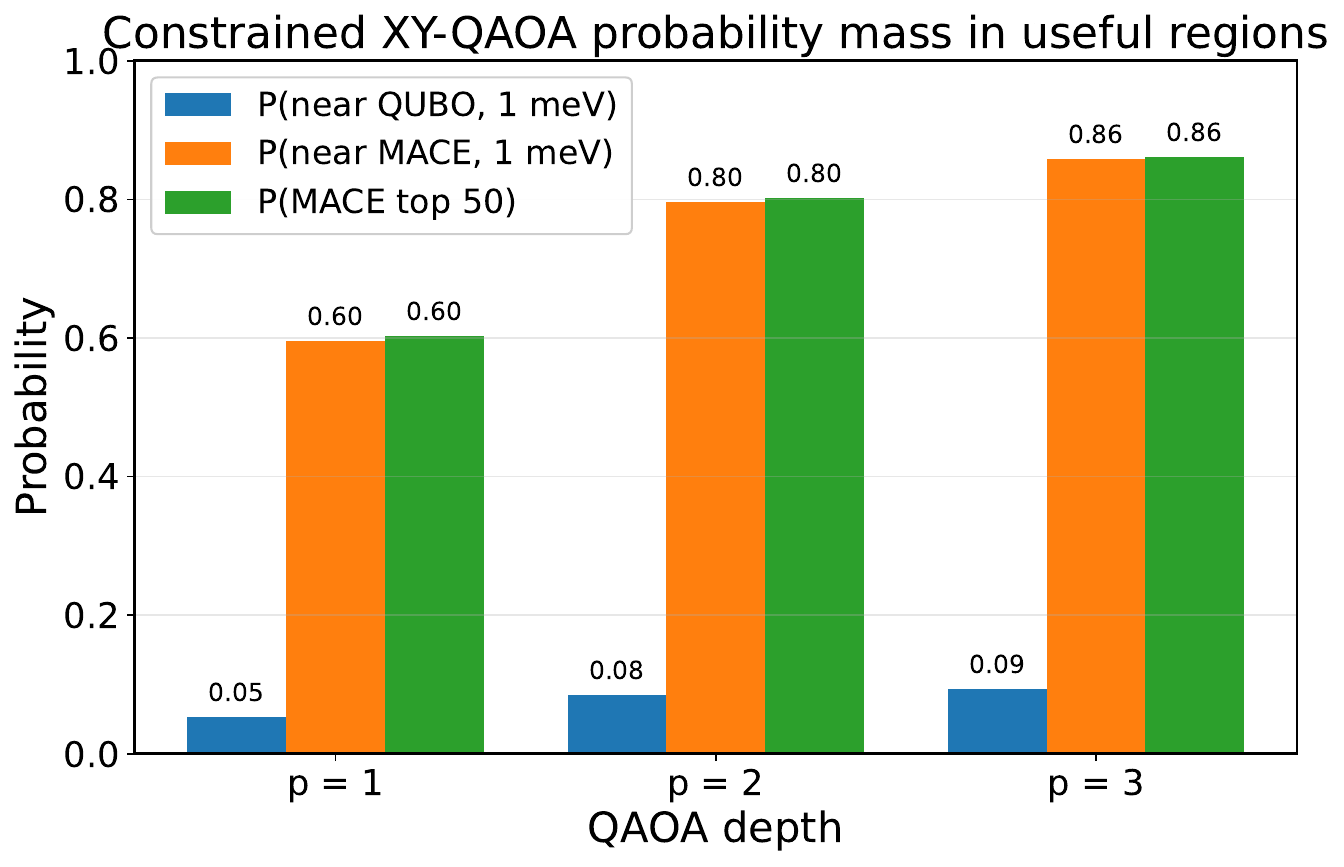}
\caption{Constraint-preserving XY-mixer QAOA: probability mass on useful
low-energy regions as a function of circuit depth $p$. All sampled states are
feasible by construction, and the probability within
\SI{1}{\milli\electronvolt} of the MACE optimum grows monotonically with depth.}
\label{fig:xyqaoa}
\end{figure}

The concentration of probability on low-energy states improves with depth. The probability of sampling within \SI{1}{\milli\electronvolt} of the MACE optimum increases from $0.596$ at $p=1$ to $0.797$ at $p=2$ and $0.859$ at $p=3$. Likewise, the probability mass assigned to the MACE top-50 configurations reaches $0.861$ at $p=3$ (Table~\ref{tab:qaoa}, Fig.~\ref{fig:xyqaoa}). This demonstrates the near-term version of the central principle of this paper: when the feasible space is much smaller than the full occupation space, encoding the constraint in the quantum dynamics is more effective than penalizing violations after the fact. The same principle motivates the feasible-space amplification discussion for fault-tolerant Grover Adaptive Search in Sec.~\ref{sec:iteration}.

\begin{table}[h]
\centering
\caption{Constraint-aware versus penalty QAOA on the Gd:ZrO\textsubscript{2}
QUBO. The XY mixer guarantees feasibility and concentrates probability on
near-optimal states; the penalty encoding does neither reliably. Probabilities
are ideal (noiseless) simulation values; ``near MACE'' means within
\SI{1}{\milli\electronvolt} of the MACE optimum.}
\label{tab:qaoa}
\begin{adjustbox}{max width=\linewidth}
\begin{tabular}{lccccc}
\toprule
Method & Depth $p$ & Feasible prob. & $P(\text{near QUBO})$ & $P(\text{near MACE})$ & $P(\text{MACE top-50})$ \\
\midrule
Penalty QAOA (best feasible) & 1 & $0.070$ & --- & --- & --- \\
Penalty QAOA (6/12 cases) & --- & $0.000$ & --- & --- & --- \\
\midrule
XY-mixer QAOA & 1 & $1.000$ & $0.053$ & $0.596$ & $0.602$ \\
XY-mixer QAOA & 2 & $1.000$ & $0.084$ & $0.797$ & $0.802$ \\
XY-mixer QAOA & 3 & $1.000$ & $0.094$ & $0.859$ & $0.861$ \\
\bottomrule
\end{tabular}
\end{adjustbox}
\end{table}

\paragraph{Logical resources and noise sensitivity.} In the 24-qubit occupation encoding, one XY-QAOA layer contains 156 two-qubit cost-phase gates from the quadratic QUBO terms and 148 two-qubit mixer interactions from the complete cation and vacancy XY-mixer graphs, for a total of 304 two-qubit gates per layer. Under all-to-all connectivity, the compiled two-qubit depth is 38 per layer and scales linearly with $p$; for example, the $p=3$ circuit uses 912 two-qubit gates and has two-qubit depth 114.

\begin{figure}[h]
\centering
\includegraphics[width=0.72\linewidth]{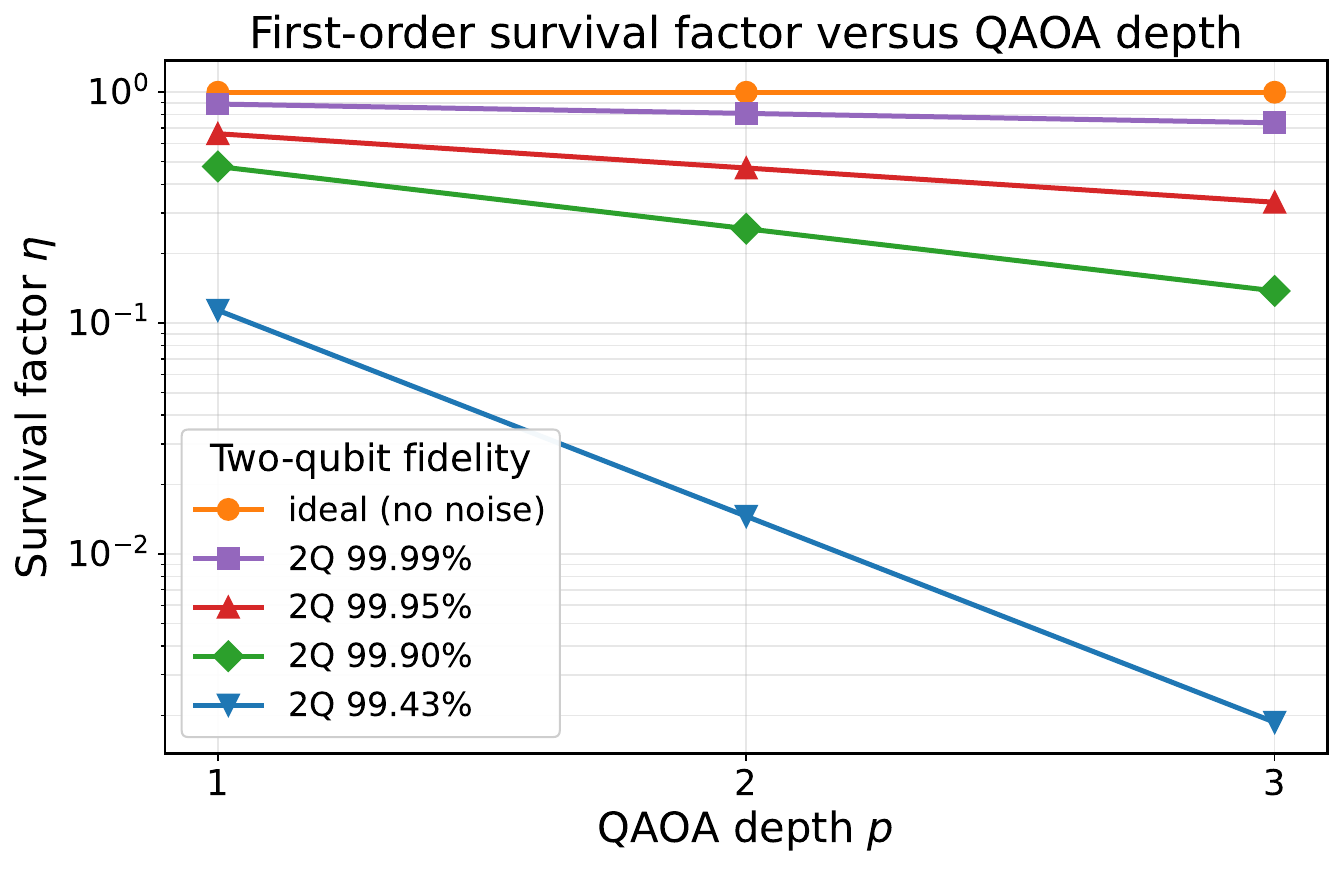}
\caption{First-order noise survival factor $\eta$ versus QAOA depth under
several two-qubit-fidelity profiles. The two-qubit gate count dominates the
error budget, producing a depth--noise trade-off between ideal concentration
and noisy survival.}
\label{fig:noise}
\end{figure}

To estimate near-term noise sensitivity, we use a first-order survival-factor
screen,
\begin{equation}
\eta = F_{1q}^{N_{1q}} F_{2q}^{N_{2q}} F_{\mathrm{ro}}^{N_q},
\end{equation}
where $F_{1q}$, $F_{2q}$, and $F_{\mathrm{ro}}$ denote the assumed one-qubit, two-qubit, and readout fidelities, and $N_{1q}$, $N_{2q}$, and $N_q$ are the corresponding operation counts. This model is not a full backend simulation; it is a screening estimate intended to identify the dominant error channel and compare depths on equal footing. As expected, the two-qubit gate count is the dominant driver of survival loss. The result is a depth--noise trade-off: deeper circuits concentrate more ideal probability on low-energy configurations, but they also accumulate more two-qubit error. At a two-qubit fidelity of \SI{99.99}{\percent}, the $p=2$ circuit retains a noise-screened near-MACE probability of $0.66$ (Fig.~\ref{fig:noise}); at \SI{99.9}{\percent}, the shallower $p=1$ circuit is preferred. These estimates should be interpreted as hardware-planning guidance rather than device-level performance predictions, and should be rerun with backend-specific calibration, routing, and noise data before experimental deployment.

% ===========================================================================
\section{GAS threshold-oracle formulation}
\label{sec:gas}

Grover Adaptive Search (GAS) recasts discrete optimization as a sequence of thresholded search problems, combining Grover amplitude amplification with the minimum-finding strategy of D\"urr and H\o yer~\citep{Grover1996,Boyer1998,Durr1996,Gilliam2021GAS}. For a threshold $T$, the oracle marks exactly those occupation bitstrings that are both feasible and lower in energy than the current incumbent,
\begin{equation}
\chi_T(z) =
\mathbf{1}\!\left[
\mathrm{feasible}(z) \wedge E_{\mathrm{QUBO}}(z) < T
\right],
\end{equation}
and the corresponding phase oracle acts as
\begin{equation}
O_T |z\rangle = (-1)^{\chi_T(z)} |z\rangle .
\end{equation}
The adaptive loop amplifies states below the current threshold, measures a candidate configuration, and tightens the threshold whenever a lower-energy candidate is found (Algorithm~\ref{alg:gas}). In this work, the central object is not only the abstract oracle $O_T$, but its explicit reversible implementation for the materials-derived QUBO.

\begin{algorithm}[h]
\caption{Adaptive GAS loop for the materials QUBO.}
\label{alg:gas}
\begin{enumerate}
\item Initialize an incumbent threshold $T$ (and $E_0$ for a random feasible $z$).
\item Build or select the constrained phase oracle
  $O_T\,|z\rangle = (-1)^{[\mathrm{feasible}(z)\,\wedge\,E_{\mathrm{QUBO}}(z)<T]}\,|z\rangle$.
\item Prepare a search state over candidate occupation configurations.
\item Apply amplitude amplification with $O_T$ and a reflection operator.
\item Measure a candidate $z$.
\item If $z$ is feasible and $E_{\mathrm{QUBO}}(z)<T$, set $T \leftarrow E_{\mathrm{QUBO}}(z)$.
\item Repeat until the threshold no longer improves or a stopping rule fires.
\end{enumerate}
\end{algorithm}

The threshold cases used in this work are controlled snapshots of the adaptive loop rather than a stochastic run of the full algorithm. They span the relevant search regimes of GAS (Fig.~\ref{fig:landscape}, Table~\ref{tab:thresholds}): a negative-control threshold below the global minimum, the ideal final threshold that marks only the optimum, small and intermediate low-energy basins corresponding to the top 5\% and top 10\% of feasible configurations, a median threshold representing an early broad-search stage, and a positive control that marks all feasible configurations. The marked-state counts are computed exactly from classical enumeration over the 448 feasible configurations and are used as validation targets for the reversible oracle.

\begin{table}[t]
\centering
\caption{Adaptive-GAS threshold ladder. Each case is a controlled snapshot of
the threshold-search loop, with exact marked-state counts over the 448
feasible configurations.}
\label{tab:thresholds}
\begin{adjustbox}{max width=\linewidth}
\begin{tabular}{lccl}
\toprule
Threshold case & Marked $M$ & Marked frac. & Role in adaptive search \\
\midrule
below\_global\_minimum     & 0   & $0.000$ & Negative control (no state can satisfy $T$) \\
global\_minimum\_only      & 1   & $0.002$ & Ideal final threshold (only the optimum) \\
top\_5\_percent            & 23  & $0.051$ & Late search: small low-energy basin \\
top\_10\_percent           & 45  & $0.100$ & Intermediate: broader low-energy basin \\
median\_threshold          & 224 & $0.500$ & Early broad search (no amplification gain) \\
above\_global\_maximum     & 448 & $1.000$ & Positive control (all states marked) \\
\bottomrule
\end{tabular}
\end{adjustbox}
\end{table}

% ===========================================================================
\section{Fixed-point encoding and reversible arithmetic oracle}
\label{sec:arith}

A reversible threshold oracle cannot compare floating-point energies directly. We therefore convert the QUBO coefficients and thresholds to fixed-point integers before circuit construction, following the standard strategy of implementing classical arithmetic reversibly with quantum networks and ripple-carry primitives~\citep{Vedral1996Arithmetic,Cuccaro2004Adder,Barenco1995Elementary,NielsenChuang2010}. For each real-valued coefficient $a_r$ and threshold $T$, we define
\begin{equation}
\tilde{a}_r = \mathrm{round}(2^{b} a_r), \qquad
\tilde{T} = \mathrm{round}(2^{b} T),
\end{equation}
with $b=18$ fractional bits. The resulting integer-valued QUBO is accumulated in a 29-bit signed range. This precision was chosen empirically to preserve all threshold labels in Table~\ref{tab:thresholds}: after quantization, the fixed-point oracle classification agrees with exact floating-point enumeration with zero mismatches.
\begin{equation}
\tilde{E}_{\mathrm{QUBO}}(z) = \tilde{E}_0 + \sum_r \tilde{a}_r f_r(z),
\qquad f_r(z)\in\{0,1\},
\end{equation}
where the feature flags $f_r(z)$ include the linear occupation variables and the quadratic products appearing in the fitted QUBO. To avoid signed reversible comparison, the arithmetic separates positive and negative weights:
\begin{equation}
P(z) = \sum_{\tilde{a}_r>0} \tilde{a}_r f_r(z),
\qquad
N(z) = \sum_{\tilde{a}_r<0} |\tilde{a}_r| f_r(z).
\end{equation}
The threshold test $\tilde{E}_{\mathrm{QUBO}}(z)<\tilde{T}$ is then rewritten as an unsigned branch-safe comparison. Defining
\begin{equation}
D = \tilde{T}-\tilde{E}_0 ,
\end{equation}
the marking condition becomes
\begin{equation}
P(z)-N(z) < D .
\end{equation}
If $D<0$, this is implemented as
\begin{equation}
P(z)+|D| < N(z),
\end{equation}
whereas if $D\ge 0$, it is implemented as
\begin{equation}
P(z) < N(z)+D .
\end{equation}
Thus, every threshold case is reduced to a comparison between two nonnegative integer registers. The comparison is carried out with a ripple-borrow comparator, after which the comparison bit is combined with the feasibility flag to produce the marked-state bit.

The full oracle construction consists of six reversible stages. First, the occupation bitstring is mapped to feature flags for the linear and quadratic QUBO terms. Second, positive and negative weighted terms are accumulated into separate integer registers. Third, the appropriate constant offset is added to the positive or negative side according to the sign of $D$. Fourth, a ripple-borrow comparator evaluates the branch-safe inequality. Fifth, the comparison result is combined with the fixed-composition feasibility check to produce the bit-oracle flag. Sixth, all arithmetic work registers are uncomputed, leaving only the occupation register and the phase kickback. Each subcomponent was built, serialized as a reloadable circuit artifact, and validated before assembly of the full constrained phase oracle.

% ===========================================================================
\section{Constrained bit oracle, phase oracle, and resource estimate}
\label{sec:oracle}

\begin{figure}[h]
\centering
\includegraphics[width=0.95\linewidth]{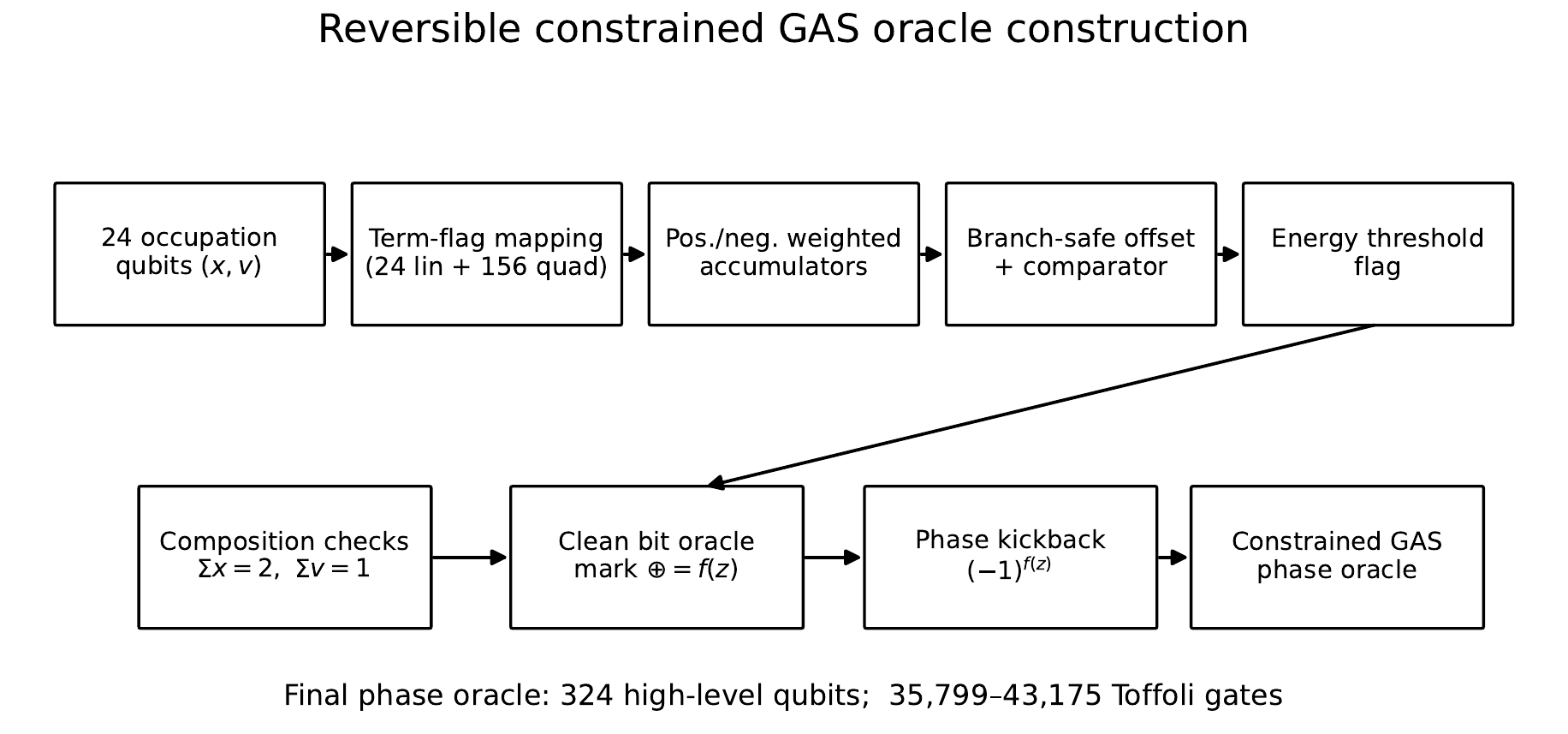}
\caption{Construction of the constrained GAS phase oracle from occupation
variables: term-flag mapping, weighted fixed-point arithmetic, feasibility
checks, clean bit-oracle marking, and phase-kickback conversion.}
\label{fig:oracle}
\end{figure}

The threshold-comparison flag alone is not sufficient for the materials search: the GAS oracle must mark only configurations that are both energetically below threshold and valid under the fixed-composition constraints. The constrained bit oracle therefore computes three Boolean flags,
\begin{equation}
x_{\mathrm{feas}} =
\left[\sum_{i=1}^{8} x_i = 2\right], \qquad
v_{\mathrm{feas}} =
\left[\sum_{j=1}^{16} v_j = 1\right],
\end{equation}
and
\begin{equation}
E_{lt} =
\left[E_{\mathrm{QUBO}}(z) < T\right].
\end{equation}
These flags are combined into a clean marker qubit according to
\begin{equation}
\mathrm{GAS}_{\mathrm{mark}}
\mathrel{\oplus}=
x_{\mathrm{feas}}
\wedge
v_{\mathrm{feas}}
\wedge
E_{lt}.
\end{equation}
All arithmetic, feasibility-checking, and temporary work registers are then uncomputed, leaving a clean reversible bit oracle. The bit oracle is converted to a phase oracle by preparing the marker in a phase-kickback state, so that marked configurations acquire a sign flip while unmarked configurations are unchanged~\citep{Grover1996,Barenco1995Elementary,NielsenChuang2010}. The resulting constrained phase oracle implements
\begin{equation}
O_T |z\rangle =
(-1)^{[
x_{\mathrm{feas}}
\wedge
v_{\mathrm{feas}}
\wedge
(E_{\mathrm{QUBO}}(z)<T)
]}
|z\rangle ,
\end{equation}
as required by the GAS threshold-search step (Fig.~\ref{fig:oracle}).

\paragraph{Headline resources.} The final constrained phase oracle requires \textbf{324 high-level logical qubits} across all threshold cases. Under a conservative clean-ancilla v-chain accounting for multi-controlled operations, the logical-qubit count ranges from \textbf{352 to 358}. The Toffoli cost ranges from \textbf{$3.6\text{--}4.3\times10^{4}$} gates across the threshold ladder (Table~\ref{tab:resources}). These are the validated concrete resources needed to apply constrained GAS to the present materials-informed QUBO, and they form the central fault-tolerant resource estimate of this work.

A key lesson is that the oracle cost is not determined by the number of binary variables alone. For this problem, the resource count is governed by the fixed-point precision, the number and signs of the fitted QUBO coefficients, the positive--negative accumulator structure, the branch-safe constant-offset comparison, the ripple-borrow comparator, the feasibility checks, and the multi-controlled marking logic. Thus, a high-level call to a Grover optimizer would hide the dominant implementation costs; explicit oracle construction is needed to obtain an auditable resource estimate~\citep{Gilliam2021GAS}.

\begin{table}[h]
\centering
\caption{Constrained phase-oracle resources by threshold case (selected
columns from the master resource table). ``High-level'' and ``v-chain'' are
two logical-qubit accountings; the Toffoli count is for the phase oracle.}
\label{tab:resources}
\begin{adjustbox}{max width=\linewidth}
\begin{tabular}{lccccc}
\toprule
Threshold case & Marked $M$ & Branch & High-level qubits & v-chain qubits & Toffoli \\
\midrule
below\_global\_minimum  & 0   & $D_{-}$ & 324 & 358 & \num{43175} \\
global\_minimum\_only   & 1   & $D_{-}$ & 324 & 357 & \num{41411} \\
top\_5\_percent         & 23  & $D_{-}$ & 324 & 357 & \num{39583} \\
top\_10\_percent        & 45  & $D_{-}$ & 324 & 352 & \num{35799} \\
median\_threshold       & 224 & $D_{+}$ & 324 & 355 & \num{38043} \\
above\_global\_maximum  & 448 & $D_{+}$ & 324 & 358 & \num{40115} \\
\bottomrule
\end{tabular}
\end{adjustbox}
\end{table}

% ===========================================================================
\section{Grover/GAS iteration baseline and feasible-space motivation}
\label{sec:iteration}

A one-iteration Grover/GAS baseline combines the constrained phase oracle with
a standard full-space diffuser over the 24 occupation qubits
(one iteration $=$ constrained phase oracle $+$ 24-bit diffuser). The diffuser
adds 43 Toffoli gates, giving a one-iteration range of
$3.6\text{--}4.3\times10^{4}$ Toffoli gates.

\begin{figure}[h]
\centering
\includegraphics[width=0.85\linewidth]{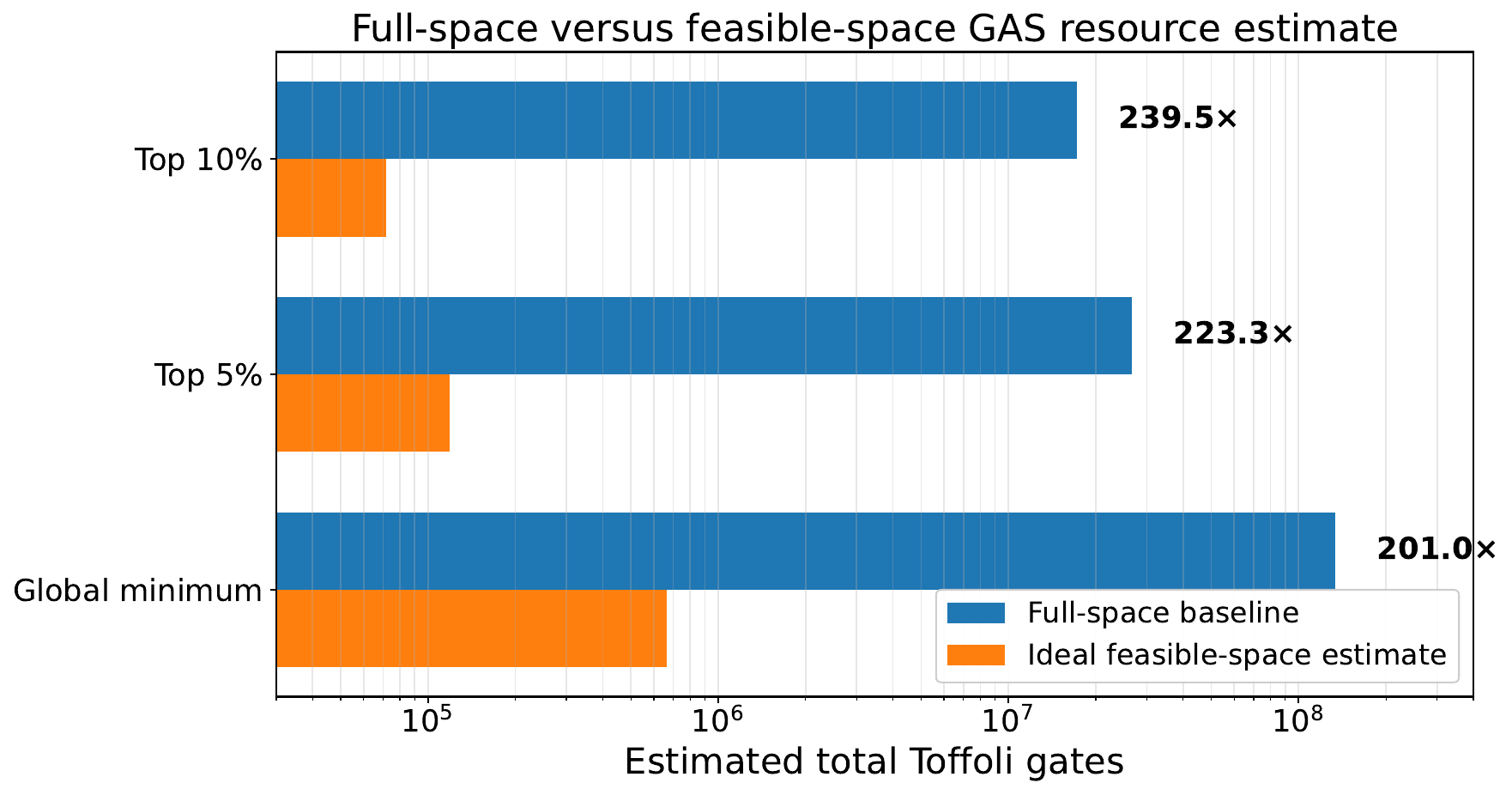}
\caption{Concrete full-space Grover/GAS baseline versus the idealized
feasible-space amplification estimate (log scale). The reduction reflects the
$\sqrt{N}$ Grover scaling between a $2^{24}$ and a 448-state search and presumes
a constraint-preserving diffuser not yet implemented.}
\label{fig:fullvsfeasible}
\end{figure}

\paragraph{Why constraint-aware amplification matters --- stated as an upper
bound.} The full-space diffuser reflects about the uniform superposition over
all $2^{24}$ bitstrings, while the physically relevant feasible subspace holds
only 448 configurations. Using the standard near-optimal iteration count
$r \approx \tfrac{\pi}{4}\sqrt{N/M}$~\citep{Boyer1998}, the iteration count for
a hypothetical feasible-space search ($N = 448$) is far smaller than for the
full space ($N = 2^{24}$). Because the per-iteration cost cancels in the ratio,
the resulting total-Toffoli reduction is essentially
$\sqrt{N_{\mathrm{full}}/N_{\mathrm{feasible}}} = \sqrt{2^{24}/448} \approx 193$,
appearing as factors of $\sim$201$\times$ (global-minimum-only) and
$\sim$240$\times$ (top-10\%) once iteration counts are rounded to integers
(Fig.~\ref{fig:fullvsfeasible}). We emphasize that this is an
\emph{idealized upper bound}: it reuses the full-space per-iteration cost and
presumes a constraint-preserving diffuser or feasible-state
preparation/reflection that we have not yet implemented. A genuine
feasible-space diffuser would in general cost more per iteration, so the
realized reduction would be smaller. The figure therefore expresses the
\emph{target} benefit of constraint-aware amplification --- the same principle
the XY-mixer QAOA already realizes on the near-term side --- rather than an
achieved fault-tolerant speedup.

% ===========================================================================
\section{Discussion}
\label{sec:discussion}

The results cohere around a single theme: for a fixed-composition materials QUBO, constraint awareness is the decisive design choice on both the near-term and fault-tolerant sides. On the near-term side, the constraint-preserving XY mixer turns a feasibility failure---penalty QAOA produces zero feasible samples in half of the tested cases---into a constraint-guaranteed search that places \SI{86}{\percent} of its probability within \SI{1}{\milli\electronvolt} of the MACE optimum. On the fault-tolerant side, the validated constrained GAS oracle quantifies the cost of marking feasible low-energy configurations, while the full-space versus feasible-space comparison shows why future amplification should be restricted to the 448-state feasible subspace rather than the full $2^{24}$ occupation space. Exact enumeration ties the two pathways together by providing a common reference for the QUBO fit, the QAOA sampling distributions, the marked-state counts, and the oracle truth tables.

Two limitations bound the claims. First, the implemented Grover/GAS iteration uses a full-space diffuser; the feasible-space reduction is therefore an upper bound until a constraint-preserving diffuser or feasible-state preparation/reflection is built. Second, the QUBO is an energy-landscape surrogate. Its aggregate fidelity is high ($R^2=0.999$), but the near-degenerate low-energy manifold means that exact identification of the MACE ground state should rely on a final MACE re-ranking of quantum-sampled candidates. This hybrid refinement is inexpensive for the present XY-QAOA results because \SI{86}{\percent} of the $p=3$ probability mass already lies in the MACE top-50 configurations.

\paragraph{Scaling outlook.}
The single 448-state instance is classically trivial by design; its value is as
a fully validated calibration point for a materials-informed quantum
optimization workflow. The oracle cost is dominated by fixed-point arithmetic
rather than by the 24 occupation qubits themselves. More generally, and
consistent with prior GAS/QUBO oracle analyses, let $n$ be the number of binary
occupation variables, $m$ the number of nonzero linear and quadratic QUBO
features that must be evaluated, and $b$ the accumulator width needed to
represent the fixed-point energy range and threshold
resolution~\citep{Gilliam2021GAS,Nagy2023FixedPointGAS}. In a straightforward
term-flag and ripple-arithmetic implementation, each nonzero weighted feature
contributes a controlled fixed-point addition whose Toffoli cost is linear in
$b$. Because the arithmetic must be uncomputed after marking, the leading
oracle cost scales as
\begin{equation}
C_{\mathrm{Toffoli}}^{\mathrm{oracle}}
=
\mathcal{O}(m b)
+
\mathcal{O}(b)
+
C_{\mathrm{feas}}(n),
\end{equation}
where the $\mathcal{O}(m b)$ term comes from weighted accumulation, the
$\mathcal{O}(b)$ term from threshold comparison and offset handling, and
$C_{\mathrm{feas}}(n)$ from the fixed-composition checks and final
multi-controlled marking logic. In the present implementation,
$n=24$, $m=180$, and $b=29$, giving validated constrained-phase-oracle costs of
$3.6\text{--}4.3\times10^{4}$ Toffoli gates. This implies a calibration scale
of roughly
\begin{equation}
\frac{C_{\mathrm{Toffoli}}^{\mathrm{oracle}}}{m b}
\approx
7\text{--}8
\end{equation}
Toffoli gates per QUBO-feature bit after including compute, comparison,
feasibility checking, marking, phase kickback, and uncomputation.

The logical-qubit scaling depends on whether QUBO feature flags are stored
simultaneously or generated and uncomputed in a streaming fashion. The present
artifact stores explicit feature and work registers, so a conservative scaling
model is
\begin{equation}
Q_{\mathrm{logical}}
=
\mathcal{O}(n + m + b + a_{\mathrm{mcx}} + a_{\mathrm{feas}}),
\end{equation}
where $a_{\mathrm{mcx}}$ denotes ancillae used for multi-controlled operations
and $a_{\mathrm{feas}}$ denotes registers used by the feasibility checks. This
explains why the present 24-variable instance requires 324 high-level logical
qubits: the dominant footprint is not the occupation register but the
materialized QUBO-term flags, fixed-point accumulators, comparator workspace,
and clean-ancilla accounting. A more space-efficient oracle could reduce the
$m$-dependent qubit footprint by recomputing or streaming term flags, but this
would generally trade qubits for additional depth or Toffoli count.

These scaling relations clarify how larger materials instances should be
interpreted. For a dense QUBO, $m=\mathcal{O}(n^2)$, so the leading Toffoli
cost scales as $\mathcal{O}(n^2 b)$; for a sparse or locality-truncated QUBO,
$m=\mathcal{O}(n)$, so the leading arithmetic cost scales only as
$\mathcal{O}(n b)$. If the fixed-point width grows only logarithmically with
the energy range and target resolution, the arithmetic overhead remains
moderate compared with the growth in the number of retained QUBO terms. Thus,
the present instance should be viewed as a calibration anchor: it gives a
validated constant-factor estimate for one concrete materials QUBO, while the
dominant route to scaling is to preserve locality or sparsity in the fitted
objective, avoid unnecessary stored term flags, and implement a
constraint-preserving diffuser so that Grover/GAS amplification acts directly
within the feasible subspace.

Future directions therefore include variable-composition QUBOs, multi-element
rare-earth screening, higher-order or sparse-local surrogate models,
constraint-preserving GAS diffusers, and multi-objective materials design
incorporating thermal conductivity, diffusivity, phase stability, and fracture
toughness. In the longer term, the same workflow can support a hybrid design
loop in which quantum optimization proposes low-energy defect configurations,
machine-learned interatomic potentials rapidly re-rank and relax the sampled
candidates, and selected structures are refined with first-principles or
property-specific simulations.

% ===========================================================================
\section{Conclusion}
\label{sec:conclusion}

We have presented an end-to-end, constraint-aware quantum optimization workflow
for fixed-composition defect-configuration search in Gd-doped
ZrO\textsubscript{2}, a representative thermal-barrier-coating material system,
and have used exact classical enumeration over the 448 feasible configurations
as a common validation reference throughout. From a MACE-MPA-0 energy dataset we
fitted a 24-variable QUBO over eight cation-occupation and sixteen
oxygen-vacancy variables. The surrogate reproduces the MACE single-point
energies with full-data $R^2 = 0.999$ and held-out $R^2 = 0.997$, and we showed
that its aggregate accuracy must be read alongside the near-degeneracy of the
low-energy manifold: a final MACE re-ranking of quantum-sampled candidates, not
the raw QUBO minimum, is the reliable route to the true ground state.

The central finding is that constraint awareness is the decisive design choice
on both sides of the quantum optimization stack. On the near-term side, a
penalty encoding fails to prepare feasible states reliably---producing no
feasible samples in half of the tested settings---whereas a constraint-preserving
XY-mixer QAOA confines the evolution to the feasible subspace by construction
and concentrates \SI{86}{\percent} of its probability within
\SI{1}{\milli\electronvolt} of the MACE optimum at depth $p=3$, with a
first-order noise screen identifying the two-qubit gate count as the dominant
error channel. On the fault-tolerant side, we did not treat Grover Adaptive
Search as a black box: we constructed and validated, layer by layer, the
fixed-point arithmetic, branch-safe comparison, feasibility checks, and phase
kickback of a constrained GAS oracle, and resource-estimated it at 324
high-level logical qubits (352--358 under conservative clean-ancilla v-chain
accounting) and $3.6\text{--}4.3\times10^{4}$ Toffoli gates per iteration. This
explicit accounting shows that the oracle cost is governed by fixed-point
precision and QUBO-term structure rather than by the bare variable count, and it
yields a calibration estimate of roughly seven to eight Toffoli gates per
QUBO-feature bit.

The contrast between the $2^{24}$ occupation space and the 448-state feasible
subspace motivates---strictly as an upper bound---a $\sqrt{N}$ feasible-space
amplification benefit of up to $\sim$240$\times$, which would become operational
only once a constraint-preserving diffuser or feasible-state preparation is
implemented. Closing that gap, alongside variable-composition and sparse-local
surrogate models, multi-element rare-earth screening, and multi-objective
materials design, is the natural next step. More broadly, the workflow
establishes a validated resource-estimation bridge from materials-informed QUBO
modeling to both variational and fault-tolerant quantum optimization, and
points toward a hybrid design loop in which quantum sampling proposes low-energy
defect configurations and machine-learned interatomic potentials rapidly re-rank
and relax the resulting candidates.

% ===========================================================================
\section*{Data and code availability}
Upon request, the authors will gladly provide the MACE energy dataset, fitted QUBO coefficients, QAOA result tables, validated oracle circuit artifacts, and the master resource table.

\section*{Acknowledgments}
The author H.S. thanks Prof.~Sanjubala Sahoo (Materials Science \&
Engineering, University of Connecticut), Prof.~Sanguthevar Rajasekaran
(School of Computing, University of Connecticut), and Dr.~Sanjeev~K.\
Nayak (Materials Science \& Engineering, University of Connecticut) for
valuable discussions during the proposal stage of this project. This work was support by internal Strategy Initiative funding and computing resources provided by P\&W, RTX.

% ===========================================================================
\bibliographystyle{unsrtnat}
\bibliography{references}

\end{document}